\documentstyle[12pt,aaspp4]{article}
\begin{document}
\title{The formation of cosmic structure with modified Newtonian
dynamics}
\author{R.H. Sanders}
\affil{Center for Astrophysics and Space Science}
\affil{University of California, San Diego}
\affil{and}
\affil{Kapteyn Astronomical Institute, Groningen, The Netherlands}

\begin{abstract}

I consider the growth of inhomogeneities in a low-density 
baryonic, vacuum energy-dominated universe in the context of 
modified Newtonian dynamics (MOND).  
I first write down a two-field Langrangian-based theory of
MOND (non-relativistic), which embodies several assumptions such
as constancy of the MOND acceleration parameter, association of a MOND 
force with peculiar accelerations only, and the deceleration of the Hubble
flow as a background field which influences the dynamics
of a finite size region.  In the context of this theory, the equation
for the evolution of spherically symmetric over-densities is non-linear 
and implies very
rapid growth even in a low-density background, particularly at the
epoch when the putative cosmological constant begins to dominate
the Hubble expansion.  Small comoving scales enter
the MOND regime earlier than larger scales and therefore evolve to 
large over-densities sooner.  Taking the initial 
COBE-normalized power spectrum provided by CMBFAST (Seljak \& Zeldarriaga
1996), I find that the final power-spectrum resembles that of the
standard $\Lambda$CDM universe and thus retains the empirical successes
of that model. 

\end{abstract}


\section{Introduction}

A primary motivation for cosmic non-baryonic dark matter          
with negligible pressure is the necessity of forming the presently
observed structure in the Universe without violating the constraints           
on temperature fluctuations in the CMB.  Basically, this is because
structure in the dark matter component on galaxy to super-cluster scales 
can begin growing 
via gravitational instability considerably before hydrogen recombination
(Peebles 1982, Vittorio \& Silk 1984, Bond \& Efstathiou 1984).  This
remains one of the powerful arguments against a low density baryonic 
universe.  Any alternative cosmology not including CDM must invoke some 
mechanism other than conventional gravitational collapse in order to form 
structure.  McGaugh (1999)
suggests that the modified Newtonian dynamics (MOND),
proposed by Milgrom (1983) as an alternative to dark matter on
on galaxy and cluster scales, can provide the needed mechanism and, further, 
that the consistency of the observed angular structure of the 
temperature fluctuations in the CMB (Lange et al. 2000, Hanany et al. 2000),
with a pure baryonic universe (McGaugh 2000), 
may be viewed as support for MOND.
This speculation is based upon the general expectation that
MOND, in providing stronger effective gravity in the limit of low 
accelerations, would assist in structure formation.

MOND is an {\it ad hoc} modification of Newton's law of inertia or gravity
at low acceleration.  The original idea is contained in the statement that
when the acceleration falls below $a_o$, a new physical constant with units
of acceleration, then the
effective gravitational acceleration approaches $\sqrt{g_na_o}$ where $g_n$ is
the usual Newtonian gravitational acceleration.  Although this simple formula
works remarkably well in describing galaxy rotation curves consistently
with the observed distribution of detectable matter (Sanders 1996, McGaugh
and de Blok 1998), it clearly 
lacks the generality to treat the problem of cosmological density
fluctuations.  

A more consistent physical description of modified dynamics is provided by  
the non-relativistic Langrangian-based theory of Bekenstein and Milgrom
(1984, hereafter BM).   An obvious procedure, when treating the growth of
density fluctuations, would be to take the modified Poisson equation of
BM and consider small fluctuations about a zeroth order solution as in 
Newtonian cosmology.  The problem is that, when applied to a finite sphere as
is usual in Newtonian cosmology, the zeroth order solution is not that of a
linear Hubble flow-- the absolute distance cannot be factored out and it is
not possible to describe cosmology in terms of a universal scale factor.  The
cosmology is basically that described by Felten (1984) and Sanders (1998), in
which MOND alters the usual Friedmann solutions; as soon as the cosmic
deceleration over some physical scale falls below $a_o$, then that  entire
region begins to deviate from uniform Hubble flow.  This leads to the eventual
re-collapse of any finite size region regardless of its original density or
expansion velocity.  In this picture density fluctuations play no role.  Apart
from problems in principle (what determines the point or points about which
MOND collapse proceeds?), this cosmology leads to clear contradictions with
observations-- recollapse in the present Universe occurs out to scales of 30
Mpc.  One might expect that in a  proper theory, the basic Hubble flow remains
intact, and structure develops from the field of small density fluctuations as
in standard gravitational collapse.

In order to construct a reasonable MOND cosmology which has this attribute,
one must supplement the BM theory with several assumptions which may
reasonably follow from a more general theory.  The first of these
assumptions-- also an aspect of the earlier MOND cosmology--  is that the MOND
acceleration parameter, $a_o$, which is comparable to the acceleration in the
outer regions of galaxies ($\approx 10^{-8}$ cm/s$^{-2}$), does not vary with
cosmic time.  Numerically, $a_o \approx cH_o/6$ which suggests that modified
dynamics may reflect the influence of cosmology on local particle dynamics. 
If $a_o$ varies as the Hubble parameter, then the argumentation presented here
would be incorrect.  However, it is also possible that $a_o$ is related to the
cosmological constant (Milgrom 1999) and is independent of cosmic time.  If
this is true then MOND plays no role in the evolution of the early 
radiation-dominated Universe since cosmic deceleration greatly 
exceeds $a_o$ on relevant
scales (e.g. the Jeans length).  In the later matter-dominated, pressureless 
evolution, the cosmic deceleration on co-moving scales corresponding to
galaxies or clusters falls below $a_o$ and one might expect modified dynamics
to affect the formation of such structure.

The second assumption directly concerns the problem of the zeroth
order Hubble flow; we wish to construct a theory in which MOND plays
no role in the absence of fluctuations, and the background cosmology is
essentially unaltered.  In other words, MOND should apply  
only to peculiar accelerations-- the accelerations 
developing from inhomogeneities-- and not to the overall Hubble flow; 
i.e., no MOND in a homogeneous Universe.  This assumption can find
some justification in the context of a stratified
scalar-tensor theory in which MOND phenomenology results from a
scalar force that becomes dominant in the limit of low scalar field
gradients (Sanders 1998).

The third assumption concerns the influence of the Hubble flow on 
the internal dynamics of an otherwise isolated spherical region. In modified
dynamics, and any covariant extention thereof, it must be the case that the
internal dynamics of a sub-system is influenced by the presence of an external
field-- the ``external field effect" (Milgrom 1983).  This is essentially an
observational requirement on MOND imposed by the absence of discrepancies in
Galactic star clusters. In other words, the underlying theory should not
respect the equivalence principle in its strong version.  With respect to
cosmology, it is not clear how the external field effect would come into
play, but I assume here that, for an over- or under-dense
spherical region, the deceleration or acceleration of the Hubble flow is the
one and only external field which influences the development of the
inhomogeneity.  Because the de-acceleration of the Hubble flow
increases linearly with scale, fluctuations on small comoving
scales are affected by MOND earlier than those on larger scale.
One might expect this to lead a hierarchical scheme of structure
formation, with smaller objects forming first.

\section{A non-relativistic Lagrangian-based theory for MOND}

It is possible to realize these three assumptions in a non-relativistic
theory of modified dynamics. Following BM I write down the Lagrangian
for a theory of MOND, as a modification of Newtonian gravity.
Although the theory is also {\it ad hoc}, this approach has several 
advantages:  As a Lagrangian-based theory it enjoys the usual properties of
conservation and consistency;
moreover, the assumptions described above are not arbitrarily inserted into 
an equation for the growth of fluctuations but are introduced at a more 
basic level.  This means that the growth equation can be derived
self-consistently from the field equations as in the Newtonian case, 
and the free parameters of the problem are evident.  Finally, the
results may actually constrain the sort of theory upon which MOND is 
to be ultimately based.

Because this is a MOND equivalent of Newtonian cosmology,
the theory described here need not be fully covariant. 
The two-field scalar theory is described by the Lagrangian
$$L_f = -{{{a_o}^2}\over{\kappa}}\Bigl[{X} 
+ (\beta X^{1\over 2} + 
{2\over 3}Y^{1\over 2})Y\Bigr]\eqno(1)$$
with $$ X = {{(\nabla\phi_1)^2}\over{a_o}^2}\eqno(2)$$
and $$ Y = {{(\nabla\phi_2)^2}\over{a_o}^2}\eqno(3)$$
where $\phi_1$ is a scalar field which we wish to identify with Newtonian
gravity and $\phi_2$ is a second field which we identify with
a MOND force.  Here $\kappa = 8\pi G$ and the fundamental acceleration 
$a_o$ is put in by 
hand but may be related to the cosmological constant in the underlying 
theory.  This is similar to the two-field version of BM theory.  The first term
($X$) is just the usual Lagrangian of a scalar field, but the second term is
anomalous and includes the MOND Lagrangian as in BM theory ($Y^{1.5}$).
However, there is an additional coupling between the two fields,
($YX^{0.5}$) where $\beta$ is a parameter, of order unity, describing the 
strength of the coupling.  

The coupling of these two fields to matter is described by the
interaction Lagrangian,
$$ L_i = -\Bigl[(\rho-2\rho_\Lambda){{\phi_1}} +
\delta\rho\phi_2\Bigr]  \eqno(4)$$
where $\rho$ is the actual density and $\delta\rho$ is the deviation
of the density from its mean value (i.e., $\rho = \sum{m_i\delta^3({\bf r}
-{\bf r_i}(t))}$; and $\delta\rho = \rho - \Pi(\tau)$ where $\Pi$ is the 
cosmological density which is only a function of cosmic time $\tau$).
The first term describes the 
coupling of Newtonian gravity to ordinary matter and the vacuum energy
density.  The second term 
describes coupling of the anomalous field to density fluctuations in the
ordinary matter and is also a 
modification of BM theory to suit the requirements of cosmology.   
The non-standard coupling to
fluctuations embodies the second assumption described in the Introduction:  
there is no anomalous MOND force in the absence of fluctuations.
This {\it ansatz}-- the most questionable step in the present procedure-- 
probably
would not be necessary in a fully covariant theory where spatial gradients
in a scalar field develop only in the presence of density gradients.

The theory is complete when we include the usual matter Lagrangian
$$L_m = \sum_i{m_i{\dot{r_i}}^2\delta^3({\bf r} - {\bf r_i}(t))}\eqno(5)$$
where $r_i$ is the position of a given particle.  Then,
the dynamics of the theory come from the action
$$S = \int{(L_f + L_i+ L_m)d^3x}.\eqno(6)$$
Taking the extremum of the action with respect
to variations in the fields, as usual, one finds the field equations
$$\nabla\cdot{\Bigl[{1\over 2}\beta a_o \hat{u}_1 
\Bigl({{\nabla\phi_2}\over{a_o}}\Bigr)^2 + \nabla\phi_1\Bigr]}
 = 4\pi G (\rho - 2\rho_\Lambda)\eqno(7)$$
$$\nabla\cdot\Bigl\{\Bigl[{{|\nabla\phi_2|}\over{a_o}} +
\beta{{|\nabla\phi_1|}\over{a_o}}\Bigr]\nabla\phi_2\Bigr\}
= 4\pi G \delta\rho.\eqno(8)$$
Here $\hat{u}_1$ is a unit vector in the direction of $\nabla\phi_1$.
Similarly, a stationary action with respect to variations in 
particle position yields
the usual particle equation of motion:
$$ {d^2{\bf{r}_i}\over {dt^2}} = -\nabla\phi_1({\bf r}_i) - \nabla\phi_2
({\bf r}_i);\eqno(9)$$
Note that this is MOND as a modification of gravity and not
of inertia.  

This is a true one-parameter theory of MOND.  One might think that the
strength of the coupling of $\phi_2$ to matter might also be adjustable
through the introduction of some additional unitless parameter,
$\alpha$ (i.e., the interaction Lagrangian would then contain
$\alpha\delta\rho\phi_2$).  But then $\alpha$ can always be absorbed
in a rescaling of the acceleration parameter $a_o$;  in fact, this 
must happen because of the requirement of MOND that the force about
a point mass asymptotically approach $\sqrt{GMa_o/r^2}$.  Therefore 
$\beta$ is the only possible free parameter of the theory; this is a 
``minimalist'' MOND theory in the sense that there are no arbitrary
functions and only one adjustable parameter.

In the limit where $\beta\rightarrow 0$, eq.\ 7 reduces to the Poisson
equation, and $\phi_1$ can be identified with the Newtonian potential.
In the absence of fluctuations ($\delta\rho = 0$),
$\nabla\phi_2=0$ and the theory becomes entirely Newtonian.  Thus
combined with  the equation of motion for a finite uniform sphere, eq.\ 7
yields the usual Friedmann equation for the time evolution of the
dimensionless  scale factor of the sphere.  Assuming, as in Newtonian
cosmology, that the scale factor of the finite sphere is identical to that of
the cosmology, the usual linear Hubble flow is recovered.  However,
in the presence of fluctuations, the second field, the MOND force,
contributes to peculiar accelerations.  Because of the coupling 
of the two fields (eq.\ 8), the zeroth order Hubble 
de-acceleration over
a finite size region appears as a background field in the determination
of the MOND peculiar accelerations (the external field effect).  

Eqs.\ 7 and 8 can be readily solved for $\nabla\phi_1$ and $\nabla\phi_2$
in the case of a spherically symmetric mass distribution representing
a bound object such as galaxy (with $\delta\rho =\rho$).  
Reasonable rotation curves 
result when $2<\beta<4$.  For smaller values of $\beta$, rotation curves
are not flat but gradually decline to the asymptotic MOND value 
($(GMa_o)^{0.25}$) and for larger
values, rotation curves rapidly decline and then rise to this asymptotic
value.

\section{The growth of fluctuations}

Here I use the notation and units from Sanders (1998):  $x$ is the
dimensionless scale factor in terms of the present scale factor; 
$\dot{x}/x$ is the Hubble parameter in terms of the
present Hubble parameter (H$_o$); and the cosmic time is in units of the
Hubble time, 1/H$_o$.  The Friedmann equation for the evolution
of the scale factor (derivable from the above theory in the absence of
fluctuations) may then be written
$$\Biggl({\dot{x}\over x}\Biggr)^2= {{\Omega_r\over{x^4}} + 
{\Omega_m\over{x^3}} - {k\over{x^2}} + {\Omega_\Lambda}}, \eqno(10)$$
where $\Omega_r$ is the present density parameter in radiation
(and other relativistic particles), $\Omega_m$ is that of ordinary 
matter (baryonic and CDM), $\Omega_\Lambda$ in vacuum energy density,
and $k=\Omega_r+\Omega_m+\Omega_\Lambda -1$ is the curvature constant.
  
In the Newtonian treatment of the development of density fluctuations, 
one may consider the time-evolution of a single Fourier component
($\delta_k$) of the fluctuation field in isolation.  This works, as long 
as the fluctuations are small, because Newtonian theory is linear;
with modified dynamics, which is fundamentally non-linear, this
is not obviously the case.  Therefore, I
consider the evolution of a sub-horizon spherical region in the Universe 
having mass $M$ and an average over (under)-density 
$\Delta = <\delta\rho/\rho>_r$.  This mean over-density
may be also identified with the mass variance over some comoving scale-- 
a quantity which is calculable from the power spectrum of Gaussian 
fluctuations (Padmanabhan 1993).  I take the total radius of the 
region to be $r_o + r_1$ where $r_o$ is the radius in 
absence of the over-density.  To first order
$$\Delta = -{{3r_1}\over{r_o}}. \eqno(11)$$
Differentiating this expression twice with respect to time we find
$$\ddot{\Delta} + 2{\dot{x}\over{x}}\dot{\Delta} + {\ddot{x}\over{x}}\Delta 
= {-3{{\ddot{r_1}}\over{r_o}}}. \eqno(12)$$
With Newtonian dynamics the peculiar acceleration would be
given by $$\ddot{r_1} = -{{2GM}\over{3{r_o}^2}}\Delta + \Omega_\Lambda
r_1 \eqno(13)$$ 
while, in the late matter and vacuum energy-dominated regime, the 
Hubble flow deceleration is 
$${\ddot{x}\over{x}} = -{GM\over{{r_o}^3}}+ \Omega_\Lambda. \eqno(14)$$
Combining eqs.\ 13 and 14 in eq.\ 12 we find the usual Newtonian expression 
for the linear evolution of fluctuations:
$$\ddot{\Delta} + 2{\dot{x}\over{x}}\dot{\Delta} = {{3\Omega_m}
\over{2x^3}}\Delta. \eqno(15)$$  

To consider how MOND may alter the growth of fluctuations, I make use of
the two-field theory described in the previous section to 
determine the peculiar acceleration in eq.\ 12 ($\ddot{r_1}$).
For the spherical mass distribution, application of the Gauss theorem
to eq.\ 7 yields
$${\beta\over 2}{{{g_2}^2}\over a_o} {\hat{u}_1}  +  g_1
= -{{4\pi G\rho}\over 3} (r_o + r_1)(1 + \Delta) + {{8\pi G\rho_\Lambda}
\over 3}(r_o+r_1). \eqno(16)$$  
where $g_1 = -d\phi_1/dr$ and $g_2 = -d\phi_2/dr$.  I set 
$g_1 = g_b + \delta g_1$ where the $g_b$ is the zeroth-order background
field given by  
$$g_b = {4\over 3} \pi G r_o (2\rho_\Lambda - \rho) \eqno(17)$$ and 
$\delta g_1$ is the small fluctuation about this background. 
Given  the zeroth order Hubble expansion and the usual definitions of
density parameters, eq.\ 17 may also be written $$g_b = -\lambda_c{H_o}^2
\Bigl[{\Omega_m\over{2x^2}} -   \Omega_\Lambda x\Bigr]\eqno(18)$$
where $\lambda_c$ is the comoving scale corresponding
to $r_o$ (i.e., $r_o = x\lambda_c$). 
  
Using eq.\ 11 to eliminate $r_1$, eq.\ 16, to first order in
$\Delta$ becomes
$$\delta g_1 + {\beta\over 2}{{{g_2}^2}\over a_o} {\hat{u}_1}
= -{{8\pi G r_o\Delta}\over 9}{(\rho + \rho_\Lambda)} \eqno(19)$$
Because of the coupling of $\phi_2$ to inhomogeneities, 
the fluctuations in the MOND field $g_2$ are always small.
Therefore, to lowest order, eq.\ 19 reduces to
$$\delta g_1 = -A\Delta\eqno(20)$$ 
where $$A = {{\lambda_c{H_o}^2}\over {3}}\Bigl[{{\Omega_m}\over {x^2}} +
\Omega_\Lambda x\Bigr] \eqno(21)$$
This is just the usual expression for the perturbed Newtonian force.

Application of the Gauss theorem to eq.\ 8 yields
$$|g_2|g_2 + \beta |g_1| g_2 = -{{4\pi a_o}G\over 3}(r_o+r_1)\Delta
= - B\Delta\eqno(22)$$
where $$B = {{a_o\Omega_m\lambda_c H_o^2}
\over{2 x^2}}\Delta\eqno(23)$$ to first order in $\Delta$.
With eq.\ 20, this reduces to
$$|g_2|g_2 + \beta g_2\Bigl|g_b-A\Delta \Bigr| = -B\Delta \eqno(24)$$ 

Thus we have two algebraic expressions for the perturbed
force fields $\delta g_1$ and $g_2$ in terms of the over- or under-density 
$\Delta$:  eq.\ 20 for the Newtonian field and eq.\ 24
for the MOND field.
We cannot immediately dismiss the
second order term in eq.\ 24.  Due to the presence of
the cosmological constant, the background field, $g_b$, will become 
vanishingly
small over some range of scale factor, and then the left-hand-side of
eq.\  24 only contains second order terms.  However, this 
quadratic equation for $g_2$ is readily solved:
$$g_2 = \pm 0.5\beta\Bigl|g_b - A\Delta\Bigr| \mp 0.5\sqrt{\beta^2(g_b -
A\Delta)^2 \pm  4B\Delta} \eqno(25)$$
where the upper sign applies to over-densities ($\Delta > 0$)  and the
lower sign to under-densities ($\Delta < 0$).

From eq.\ 9, the total acceleration of a 
spherical shell is $$\ddot{r_o} + \ddot{r_1} = g_b+\delta g_1+g_2.
\eqno(26)$$
Given that $\ddot{r_o} = g_b$ (the Friedmann equation), then
$$\ddot{r_1} = \delta g_1 + g_2.\eqno(27)$$
Substituting eqs.\ 27, 21 and 20 into eq.\ 12, the growth equation becomes
$$\ddot{\Delta} + 2{\dot{x}\over{x}}\dot{\Delta} = {{3\Omega_m}
\over{2x^3}}\Delta - {{3 g_2}\over{x\lambda_c}} \eqno(28)$$
with $g_2$ given by eq.\ 25  supplemented by eqs.\ 23, 21 and 18.  This is 
the basic equation for the evolution of spherically symmetric over-
or under-dense regions of the Universe in the context of 
the assumed two-field theory of modified
dynamics.  If $a_o\rightarrow 0$, then the final term (the MOND field $g_2$)
vanishes and the equation is identical to eq.\ 15 for the Newtonian
growth of perturbations.  The second term dominates only when
$|g_b|<a_o$.

This second order differential equation is non-linear,
even in the regime where $\Delta$ is small.  The non-linear term in eq.\ 28
(or in $g_2$ given by eq.\ 25) 
only becomes important when $g_b\rightarrow 0$, but  
this does happen within a Hubble time in a Universe having a cosmological 
constant comparable to ${H_o}^2$.  It is also noteworthy that, 
unlike the Newtonian case, the growth of an over-density depends upon the 
comoving scale.  This is primarily because smaller comoving scales
enter the MOND regime earlier; specifically, the scale factor at which
a fluctuation enters the MOND regime, given by the condition that
$|g_b|=a_o$, is $$x_c=\sqrt{{\Omega_m\lambda_c}\over{2a_o}}H_o \eqno(29)$$
in the matter-dominated regime. 

\section{Numerical results}

\subsection {The growth of inhomogeneities}

Using a fourth order Runge-Kutta technique, I have numerically integrated
eq.\ 28 for the growth of over-densities in McGaugh's (2000) vacuum energy
dominated baryonic Universe:  $\Omega_m=\Omega_b= 0.034$, 
$\Omega_\Lambda=1.01$, h = 0.75.  
The calculations are initiated at $x=1.37\times 10^{-3}$ or at a 
redshift of 730
which is roughly the epoch of matter-radiation equality
(in this low density Universe matter-radiation equality occurs after
hydrogen recombination).  In all calculations I take $a_o=1.2\times
10^{-8}$ cm s$^{-2}$ (scaled to $H_o =
75$ km s$^{-1}$ Mpc$^{-1}$) as determined by fitting to the observed
rotation curves of nearby galaxies (Begeman et al. 1991).   Smaller
values of the coupling between the Newtonian and MOND fields, $\beta$, yield
more rapid growth.  As $\beta \rightarrow 0$ the background field effect
vanishes and only the peculiar accelerations enter the MOND  equation.  In
that case the non-linear term in the growth equation always dominates and the
growth of fluctuations is much too rapid.  With respect to matching 
the form of galaxy rotation curves $\beta$ should lie between 2 and 4.
In all calculations described below
I take $\beta = 3.5$. 

Fig.\ 1 illustrates the growth of spherically symmetric over-densities  
having comoving radii
of 20, 40 and 80 Mpc.  The initial over-density is assumed to be 
$\Delta_i=2\times 10^{-5}$ which is comparable to the COBE-normalized
fluctuation amplitude on these scales in standard cosmology
(Bunn \& White 1997). The solid lines follow the
MOND evolution of $\Delta$ as a function of scale factor (eq.\ 28).  
The dotted line is the standard Newtonian evolution (eq.\ 15).  

It is evident that, in the standard model of structure formation by
gravitational instability, a region with this initial over-density
could not grow to the non-linear regime by the present epoch.
The COBE-normalized fluctuations in a baryonic universe, with
$\Omega_b$ of a few $10^{-2}$, could not possibly develop into the
observed large scale structure.  On the other hand, with 
modified dynamics, the growth becomes nearly exponential from the 
time a region enters the low acceleration regime up to the point  
where the cosmological
term dominates in the Friedmann equation.  The most rapid growth occurs
at the point where $g_b\rightarrow 0$ and the non-linear term dominates;  
thus the cosmological constant, by permitting the background cosmological
acceleration to vanish, actually promotes the rapid growth of fluctuations. 

The development of non-linear
structure on the scale of tens of Mpc would certainly seem possible.
The fact that the smaller scale fluctuations enter the non-linear
regime earlier is consistent with a hierarchical scheme (with galaxies
forming before clusters), as suggested in Sanders (1998).  However, unlike
the scenario sketched there, the underlying Hubble flow remains
intact.

\subsection{The power spectrum of fluctuations}

Through use of eq.\ 28 it is straightforward to determine 
the present power spectrum of fluctuations generated by MOND growth.
The initial COBE-normalized power spectra is provided 
by CMBFAST (Seljak \& Zeldarriaga 1996) at the 
epoch of matter-radiation equality (z = 730) in the McGaugh universe
(assuming the primordial fluctuation spectrum to be standard 
Harrison-Zeldovitch, n=1).  A measure of over- (under-) density 
on various scales is the mass-variance, ${\Delta_k}^2$, per unit 
interval in log wavenumber; this is related to the power spectrum
as $${\Delta_k}^2 = 2\pi^2 k^3 P(k) \eqno(30)$$
(Padmanabhan 1993).  I follow the evolution of these over-densities 
with amplitude $\Delta_k$ on a sub-horizon scale 
again by numerical integration of eq.\ 28.  Super-horizon growth of large 
regions is
followed crudely by considering the density evolution of sub-universes
with the curvature constant appropriate to a given initial $\Delta_k$.
When a spherical region enters the horizon, eq.\ 28 is 
immediately applied
for the subsequent growth.  This method is not precise for those
regions which enter the horizon in the matter dominated era
($z<700$) but should yield an accurate representation of the
MOND power spectrum for smaller scale fluctuations ($k>0.01$).

I convert the final over-densities back to the power
spectrum by again applying eq.\ 30.
The resulting power spectrum at the present epoch is shown by the solid
curve in Fig.\ 2.  The oscillations with wave number
are the relic of the acoustic oscillations frozen into the plasma at the 
epoch of recombination (in dark matter-dominated models these are 
suppressed).  Also shown in the same figure is the power spectrum that 
would result from the identical initial spectrum but with only Newtonian 
growth (dotted curve).  The abrupt decrease in power where $k>0.05$ is 
due to Silk damping (photon diffusion) of the fluctuations on 
scales smaller than about 20 Mpc by the epoch of recombination (Silk 1968). 

Even with Silk damping there is large amplification of
the power spectrum at these small scales due to effect of modified dynamics.
On scales smaller than 10 Mpc the structure has already become highly
non-linear ($\Delta_k>1$), and the calculated power spectrum is not to be 
trusted.  

Also shown on the same figure is the zero redshift power spectrum 
of fluctuations in the context of the standard $\Lambda$CDM model.  
This, less the oscillations, is very similar to the MOND power
spectrum on currently measurable scales ($k>0.01$).  In other words, MOND 
in a pure baryonic universe mimics quite closely the power spectrum
in this favored dark matter model on scales of 10 to 100 Mpc; i.e.,
the phenomenological successes of the standard model are retained.  
In particular, the usual 
measure of the amplitude of inhomogeneities in the present Universe, 
$\sigma_8$ is found to be 1.5 from the MOND power spectrum, which is
about 50\% larger than that of the $\Lambda$CDM model.

\subsection{Peculiar velocities}

One may also consider the predicted peculiar velocities as a test of 
this scenario (bearing in mind the obvious dangers of assumed spherical
symmetry and non-linearity
on the scale of 10 to 20 Mpc).  

The estimation of peculiar velocities follows from the continuity
equation as in the usual treatment (Peebles 1999).  Taking the first
time derivative of eq.\ 11, we find
$$\dot{r_1} = -{{r_o\dot{\Delta}}\over 3}
- {{\Delta r_o H_o}\over 3}. \eqno(31)$$
The peculiar velocity at
$r = r_o+r_1$ is then just $$ V_p = (\dot{r_o} + \dot{r_1}) - H_o(r_o+r_1),
\eqno(32)$$ which, again after use of eq.\ 11, becomes
$$V_p = - r_o\dot\Delta/3. \eqno(33)$$
On the comoving scale of the
Virgo cluster ($Hr_o \approx 1300$ km/s), the over-density of galaxies is
roughly $\Delta \approx 2$ (Strauss et al. 1992).  
Choosing an initial $\Delta$ on this comoving
scale to match the present observed over-density of galaxies,
I find from numerical integration of eq.\ 28 
that $v_p = 450$ km/s.  This is larger than present estimates,
although widely disparate values have been reported (Tonry et al. 2000).  
Although somewhat larger values of the coupling parameter $\beta$ can result
in less vigorous stirring of the local Universe, it is probably premature 
to tinker with the crude MOND theory described here.

\section{Conclusions}

Here I show that MOND provides the possibility of overcoming
the problem of structure growth in a low-density baryonic Universe.
In the context of the simple two-field non-relativistic theory of
modified dynamics presented here, we see that
when the background deceleration of the Hubble flow over a 
given scale falls below the critical MOND acceleration, 
$a_o$, then the growth of structure on that scale is
greatly enhanced relative to the Newtonian expectation.  The growth of
over-densities on smaller scale is even more enhanced due to the
fact that smaller regions enter the MOND regime earlier;  the
early growth of small-scale fluctuations can compensate for the effect of
Silk damping on these scales.  Thus the resulting power spectrum,
apart from the oscillations, closely 
resembles that of the favored $\Lambda$CDM cosmology.  On the scale of
galaxies (1.5 Mpc), even though the typical initial over-density 
is on the order of $2\times 10^{-10}$, the fluctuation grows to the 
non-linear regime by a redshift of 2.5.   Thus MOND would appear not
only to explain the observed large scale structure, but also provide
a mechanism for early galaxy formation.  This
is all achieved with the value of $a_o$ determined from galaxy rotation
curves. The minimalist MOND theory has not been
fine-tuned in any sense to match the observed power spectrum; the single
adjustable parameter $\beta$ lies within the range which is consistent
with the observed form of galaxy rotation curves.   

Of course, these conclusions depend upon the approximate validity of the 
assumed theory described in section 2.
This theory, and assumptions embodied therein, guarantee that
the early, radiation-dominated evolution of the universe is identical
to that of the standard model, that the basic Hubble
flow is unaffected by MOND, and that fluctuations on
the scale of galaxies to super-clusters enter the MOND regime, determined
by the background Hubble flow, sufficiently early (but not too early) to
assure growth to the present amplitude.  

Because of the necessity of such an {\it ad hoc} theory 
in the absence of a more fundamental covariant theory, 
it is perhaps premature to compare in detail the predicted
power spectrum or peculiar velocities with observations.  
In particular, the oscillations (Fig.\ 2) may not actually be evident in 
the evolved power spectrum due to  
non-linear aspects of the theory which are ignored here--
specifically, not only individual Fourier components 
but also over-dense spherical regions cannot be considered in 
isolation (larger scale peculiar accelerations contribute to the
background field).  The detailed results shown in Figs.\ 1 and 2 should be 
taken as a demonstration that MOND, in a low-density baryonic universe, 
can provide a vigorous growth of fluctuations-- growth which is 
sufficiently rapid to lead to the large scale structure observed
at the current epoch.

Finally, I re-emphasize that the presence of a dynamically significant 
cosmological constant plays a necessary role in the rapid growth of structure 
with this version of modified dynamics.  The MOND growth of inhomogeneities
accelerates at the epoch when $g_b\approx 0$ 
due to the dominance of the non-linear term in eq.\ 28  
(only possible with a cosmological constant comparable to ${H_o}^2$).
This adds a new aspect to
an anthropic argument originally given by Milgrom (1989):  we are 
observing the Universe at an epoch when $\Omega_\Lambda$ has only 
recently emerged as the dominant term in the Friedmann equation 
because it is only then that structure formation proceeds rapidly.

If the evidence in support of a baryonic-$\Lambda$ universe continues
with further observations of the CMB angular power spectrum,  
then some unconventional mechanism for the formation of structure must
be invoked.  Here it is evident that modified dynamics, with a 
well-documented success in explaining the kinematic observations
of galaxies and clusters without dark matter, may also successfully
address the problem of structure formation in a low-density baryonic
universe.

I am grateful to Art Wolfe, Eric Gawiser and Kim Griest
for teaching me all about anisotropies in the CMB
and the standard treatment of the evolution of density fluctuations.  
I thank Stacy McGaugh for stimulating this work, Arthur Kosowsky and
Jacob Bekenstein for helpful criticisms, and the referee, Jim Peebles,
whose numerous critical remarks led to a considerable improvement in
the content and presentation of this paper.  Finally I am most grateful
to Moti Milgrom for his typically penetrating comments on possible 
MOND cosmologies.

\newpage
\figcaption[mff1.ps]{The growth of spherically symmetric 
over-densities, $\Delta$, in a 
low-density baryonic universe as a function of scale factor in the context
of the two-field Langrangian theory of MOND.  The solid
curves correspond to regions with comoving radii of 20, 40 and 80 
Mpc.  The dotted line is the corresponding
Newtonian growth.  With MOND smaller regions enter the 
low-acceleration regime sooner and grow to larger final amplitude.
The vertical dashed line indicates the epoch at which the cosmological
constant begins to dominate the Hubble expansion in the assumed model
cosmology}

\figcaption[mff2.ps]{The present power spectrum of fluctuations in the
low density baryonic universe where the growth is determined by 
the MOND modified gravity theory (solid curve).  The initial 
amplitudes are taken from
the COBE-normalized  provided by CMBFAST at the epoch of matter-radiation
equality.  Also
shown (dotted curve) is the power spectrum which would result from Newtonian
growth of the same initial fluctuation spectrum.  The
dashed curve is the power spectrum in the standard $\Lambda$CDM cosmology. 
In all cases h=0.75.}

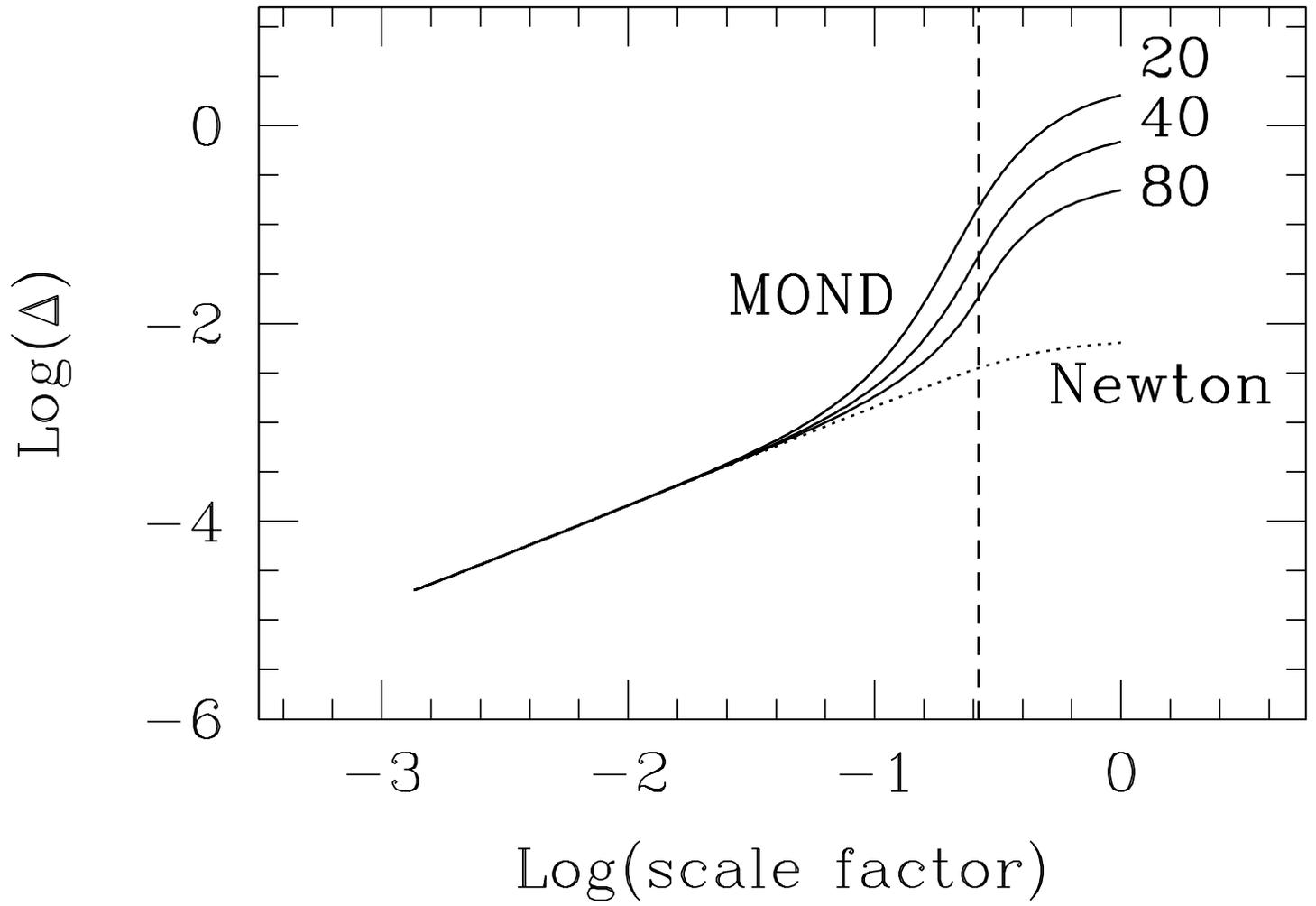
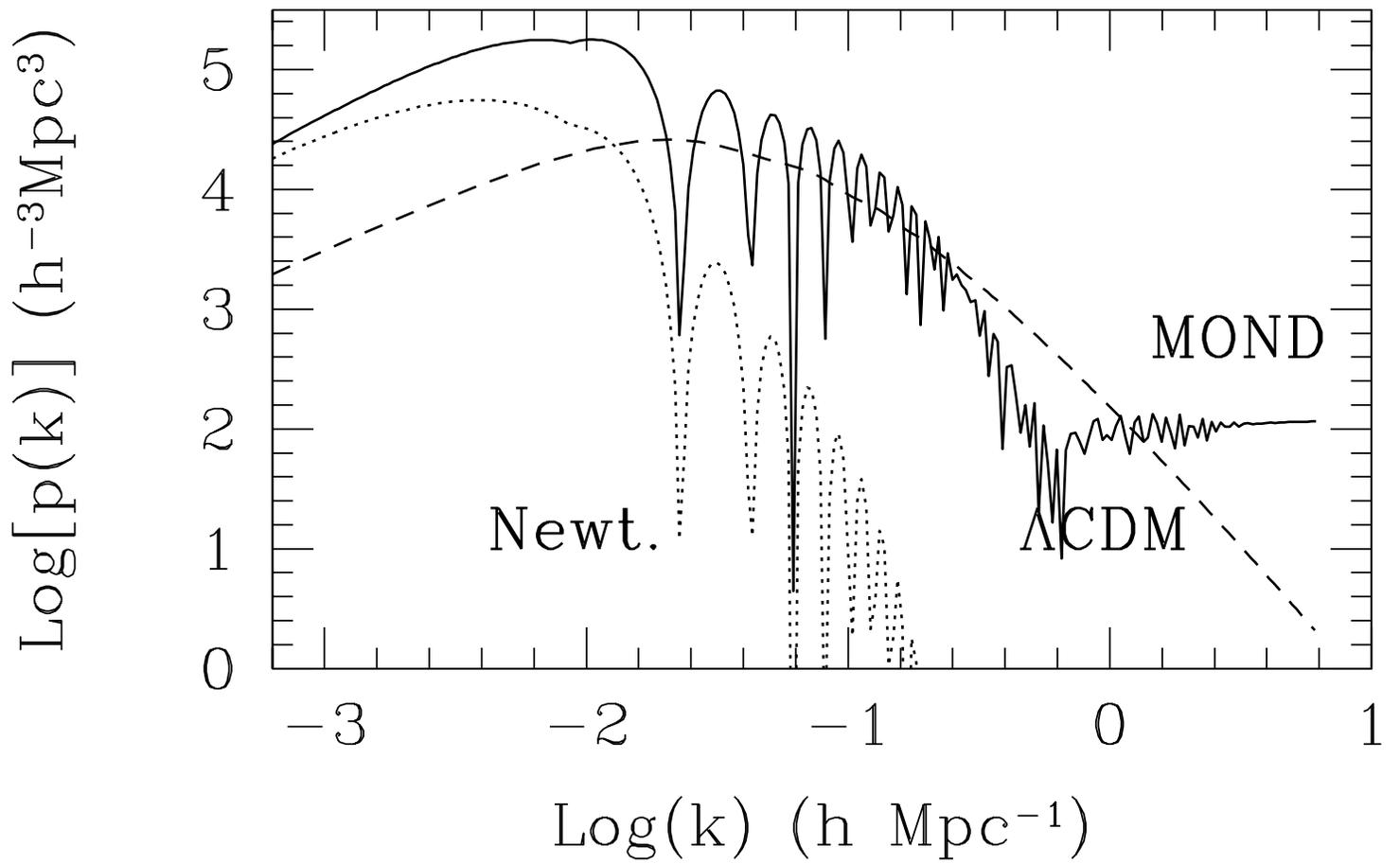
\end{document}